%% file: Letter_draft_arxiv.tex
\documentclass[journal,onecolumn]{IEEEtran}
\usepackage{amsthm,amssymb,graphicx,multirow,amsmath,color,amsfonts}%,ulem}
\usepackage[update,prepend]{epstopdf}
\usepackage[noadjust]{cite}
\usepackage[latin1]{inputenc}
\usepackage{tikz}
\usetikzlibrary{arrows,calc}		% Optional ticks libraries
\usepackage{bbm} % for \mathbbm{1}
\usepackage{pdfpages}
\usepackage{tabulary}
\usepackage{multirow}
\usepackage{comment}
\usepackage{subfigure,array}
\usepackage[linesnumbered,ruled]{algorithm2e}
\usepackage{xcolor}

\allowdisplaybreaks
\include{notation}

\let\emptyset\varnothing

\title{Meta Distribution of Downlink $\sir$ in a Poisson Cluster Process-based HetNet Model}   % type title between braces
\author{Chiranjib Saha, Mehrnaz Afshang, and Harpreet S. Dhillon       % type author(s) between braces
\thanks{C. Saha and H. S. Dhillon are with Wireless@VT, Department of ECE, Virgina Tech, Blacksburg, VA, USA. Email: \{csaha,  hdhillon\}@vt.edu. M. Afshang is with  Ericsson Research, Santa Clara, CA, USA, Email:\  mehrnaz.afshang@ericsson.com.

The support of the US National Science Foundation (Grant CNS-1617896)  is gratefully acknowledged. 
}}
\begin{document}
\maketitle
\begin{abstract}
The performance analysis of heterogeneous cellular networks (HetNets), that relied mostly on the homogeneous Poisson point process (PPP) for the spatial distribution of the users and base stations (BSs), has seen a major transition  with the emergence of the Poisson cluster process (PCP)-based models. With the combination of PPP and PCP, it is possible to construct a general HetNet model which can capture formation of hotspots and spatial coupling between the users and the BSs.  
While the downlink coverage analysis of this model in terms of the distribution of the received downlink signal-to-interference ratio ($\sir$) is well understood by now, more fine grained analysis in terms of the meta distribution of $\sir$ is an open problem. In this letter, we solve this problem by deriving the meta distribution of the downlink $\sir$ assuming that the typical user connects to the BS providing the maximum received power.
%Moreover, these models are quite tractable in terms of the analysis of the distribution of the downlink signal-to-interference ratio ($\sir$) of the typical user. With the goal of extending the analytical framework of the PCP-based general HetNet model further, in this letter, we characterize the meta distribution of the downlink $\sir$ assuming that the typical user connects to the BS providing the maximum received power. 
\end{abstract}
\section{Introduction}\label{sec::intro}
The last few years have seen two major enhancements in the baseline approach to the modeling and analysis of cellular networks using stochastic geometry. (i) {\em Enhancements in the model:} While the baseline network models relied on homogeneous  PPPs to model the spatial distribution of the BSs and users~\cite{andrews2016primer}, the  recent efforts have focused on using more sophisticated point processes  to capture the spatial couplings    between the locations of the BSs and users.  
A key set of works in this direction is based on the  PCPs which along with PPPs result in a more general HetNet model~\cite{saha2019unified} with the PPP-based baseline network model being its special case.  (ii) {\em Enhancements in the metrics:} The conventional analyses of HetNets using stochastic geometry have  focused on  the coverage probability which is the  complementary cumulative distribution function (CCDF) of the signal-to-interference-and-noise-ratio ($\sinr$). While coverage is a useful {\em first-order} metric, it does not provide any  information on the variation of $\sinr$ over the network. To obtain a more fine-gained information on the $\sinr$ performance of the network, it is important to characterize the meta distribution of $\sinr$ from which  the $\sinr$-coverage can be obtained as a  special case~\cite{metadistribution}.  While the meta distribution of $\sinr$ has been extensively studied for the {\em baseline PPP-based HetNet models}, this characterization for the PCP-driven general HetNet model, proposed in~\cite{saha2019unified}, is not known, which is the main objective of this letter.

{\em Prior Art.} 
The coverage analysis of the PPP-based cellular models is fairly mature by now (see \cite{andrews2016primer} and the references therein). The meta distribution of $\sir$ was first studied in~\cite{metadistribution} for the Poisson bipolar and cellular networks. It was subsequently extended to the $K$-tier PPP-based HetNet model in~\cite{wang2018sir,Deng_meta_rate}. On the modeling side, a more general HetNet model based on the combination of PPPs and PCPs was recently proposed in~\cite{saha2019unified,saha20173gpp,afshang2018equi}. While this model yields several spatial configurations of cellular network that are of practical interest (including the baseline PPP-based model as its special case), its analytical treatment thus far has been limited to the coverage probability. In this letter, we derive the meta distribution of the downlink $\sir$ for this model.

%The coverage analysis of the PPP-based models is quite mature by now  (see \cite{andrews2016primer,elsawy2016modeling} and the references therein). The existing framework of coverage analysis is generalized in~\cite{metadistribution} by the formulation of the meta distribution for the PPP-based network models. Following this pioneering work, the authors in \cite{wang2018sir,Deng_meta_rate} have derived meta distributions of $\sinr$  for the  $K$-tier PPP-based HetNet model.  Recently, a more general spatial model for HetNets based on the combination of PPPs and PCPs has been proposed in~\cite{saha2019unified,saha20173gpp}. Even though  this   model yields several  spatial configurations of cellular network that are of practical interest (including the baseline PPP-based model), the analytical treatment of this model is   limited to the derivation of coverage probability. In this letter, we extend this framework to characterize the meta distribution of $\sinr$.

{\em Contributions.} 
In this letter, we consider a general $K$-tier HetNet 
model where the BSs and the user locations are modeled as either a PCP or a homogeneous PPP. For this model, we  characterize the meta distribution of downlink $\sir$ of the typical user assuming that the network is operating in an  interference-limited regime and the typical user connects to the BS providing maximum received power averaged over fading. 
%We  user locations are independent of the BS locations and \case~2: where the user locations are coupled with the locations of the BSs,  which is distributed as a PCP.
To enable the analysis, we  construct an equivalent single tier cellular network  
%which will have the same distribution of $\sir$ as the 2-D $K$-tier network 
by projecting the BS point processes in $\R^2$ on the positive half line $\R^+$ that will have the same distribution of $\sir$ as the original 2-D $K$-tier network.
% \chb{This projection allows us to obtain a more compact derivation of the meta distribution of $\sir$ compared to the approach taken in \cite{saha2019unified}}. 
Although the equivalence of the analyses using this  single tier network in $\R^+$ and the $K$-tier HetNet in $\R^2$   is quite well-known for the PPP-based model (see \cite{MadhusudhananRestrepoBrown2016}), this letter makes the first attempt to develop this approach for the new PCP-based HetNet models. 
 While this alternate analytical framework for the general HetNet model is novel in its own right,  
 we use this framework to  derive the exact analytical expressions of the $b$-th moment of the conditional success
probability for the typical user  under Rayleigh
fading.   The exact expression of the meta distribution being computationally infeasible, we use the $b$-th moments to  provide  an accurate beta approximation of the CCDF of the meta distribution. 
%{\bf Notation:} We use bold-style letters (${\bf z}$) to denote vector in $\R^2$, serif-style letters ($z$) to denote Euclidean norm, i.e., $z=\|{\bf z}\|$.
\section{System model}
\label{sec::system::model}
We model a HetNet as a $K$-tier cellular network in which BSs in the  $i$-tier are distributed as a point process $\{{\bf x}\}\equiv\Phi_i\subset\R^2$  and transmit with power $P_i$, which is assumed to be fixed for all the BSs in $\Phi_i$. The point process $\Phi_i$ is either a homogenous PPP with intensity $\lambda_i$ or a  PCP. We denote the index sets of the BS tiers being modeled as PPP and PCP by ${\cal K}_1$ and ${\cal K}_2$, respectively with $|{\cal K}_1|+|{\cal K}_2|=K$. While PPP, used as a baseline spatial model for cellular networks~\cite{andrews2016primer} needs no introduction, we define PCP for  completeness as follows. 
%\chr{We may not want to have PPP BSs, check this}
\begin{ndef}\label{def::PCP} A PCP  $\Phi_i(\lambda_{{\rm p}_i}, \bar{m}_i, f_i)$ for $i\in{\cal K}_2$ is defined as:
\begin{align*}
\Phi_i(\lambda_{{\rm p}_i}, \bar{m}_i, f_i)=\bigcup\limits_{{\bf z} \in \Phi_{{\rm p}_i}}{\bf z}+{\cal B}^{\bf z}_i,
\end{align*}
where $ \Phi_{{\rm p}_i}$ is the  parent PPP with intensity $\lambda_{{\rm p}_i}$ and ${\cal B}_i^{\bf z}$ is the offspring point process. The offspring point process is a sequence of independently and identically distributed (i.i.d.)  random variables  with probability density function (PDF) $f_i({\bf s})$. The number of points in ${\cal B}_i^{\bf z}$ is Poisson distributed with mean $\bar{m}_i$.
\end{ndef}
{We further assume that   the offspring points are isotropically distributed around the cluster center. Thus the  joint PDF of the radial coordinates of the offspring points with respect to the cluster center is denoted as: $f_i(s,\theta_s) = f_{i(1)}(s)\frac{1}{2\pi}$. That said,  the PDF  of the distance of a point of $\Phi_i$ from the origin given its cluster center at ${\bf z}\in \Phi_{{\rm p}_i}$ is given by:  
$f_{{\rm d}_i}(r|{\bf z}) =f_{{\rm d}_i}(r|{z})  $, where $\|{\bf z}\| = z$. 
  For the numerical results, we choose a special case of PCP,   known as Thomas cluster process (TCP) where 
 the offspring points in ${\cal B}^{\bf z}$ are distributed normally around the origin, i.e.,
%\begin{align}\label{eq::density_thomas_definition}
$f_i({\bf s}) = \frac{1}{\sigma_i^2}\exp\left(-\frac{\|{\bf s}\|^2}{2\sigma_i^2}\right).$
 Here $\sigma_i^2$ is the cluster variance.}  
   When $\Phi_i$ is a TCP, the conditional distance distribution given $\|{\bf z}\|=z$ is Rician with PDF: 
 \begin{align}\label{eq::marginal::dist::tcp}
f_{{\rm d}_i}(x|z)
= \frac{x}{ \sigma_i^2} e^{-\frac{x^2+z^2}{2 \sigma_i^2}} I_0\left(\frac{x z}{\sigma_i^2}\right),  x,z\geq 0, i\in{\cal K}_2,
\end{align} where $I_0(\cdot)$ is the modified Bessel function of the first kind with order zero.
 We now focus on the user point process which is denoted as $\Phi_{\rm u}$. We consider two types of users in the network:
\begin{itemize}
\item { \case~1} (independent user and BS point processes): $\Phi_{\rm u}$ follows a stationary distribution independent of the BS point processes.
\item { \case~2}  (coupled user and BS point processes): $\Phi_{\rm u}$ is  a PCP with the same parent point process as that of $\Phi_q$  for some $q \in {\cal K}_2$ with cluster variance $\sigma_q^2$.
\end{itemize}
We now focus on the typical user in this network. Since the network is stationary, we can assume that the typical user is located at the origin.  It should be noted that the selection of the  typical user in \case~2  implies the selection of a cluster of $\Phi_{q}$  as well. We denote the center of this BS cluster by ${\bf z}_0$. As a consequence, $\Phi_{q}$ is always conditioned on having a cluster ${\bf z}_0+{\cal B}^{{\bf z}_0}_q$.  Thus the typical user  perceives the palm version of $\Phi_q$,  which, by Slivnyak's theorem~\cite{chiu2013stochastic} is equivalent to 
 $\Phi_{q}\cup {\bf z}_0+{\cal B}^{{\bf z}_0}_q$ where $\Phi_{q}$ and ${\bf z}_0+{\cal B}^{{\bf z}_0}_q$ are independent. For \case~1 users, this construction does not arise since the selection of the typical user does not impose any restriction on the BS distributions. In order to unify the analyses of \case~1 and \case~2 users, we define $\Phi_0$ as a set of BSs whose locations are coupled with that of the typical user as follows:
\begin{align}
\Phi_0=\begin{cases}
\emptyset;& \text{\case~1},\\
{\bf z}_0+{\cal B}^{{\bf z}_0}_q;&\text{\case~2}.
\end{cases}
\end{align}
 The BS point process perceived by the typical user can be  defined as the superposition of $K+1$ BS point processes: $\Phi=\cup_{i \in {\cal K}}\Phi_i$, where ${\cal K}={\cal K}_1\cup {\cal K}_2 \cup \{0\}$. The downlink $\sir$  of the typical  user is denoted as:
\begin{align}\label{eq::sir::def}
\sir = \frac{P_k h_{{\bf x}^*} \|{\bf x}^*\|^{-\alpha}}{\sum_{i\in {\cal K}}\sum_{{\bf x}\in{{\Phi}_i}\setminus\{{\bf x}^*\}} P_i h_{{\bf x}} \|{\bf x}\|^{-\alpha} },
\end{align}
where $\{h_{{\bf x}}\}$ is an i.i.d. sequence of random variables where $h_{\bf x}$ is the fading coefficient  associated with the link between the typical user and the  BS  at ${\bf x}$. We assume Rayleigh fading i.e. $h_{\bf x}\sim \exp(1)$  and $\alpha>2$ is the path loss exponent. Here $\|{\bf x}^*\|$ is the location of the serving BS which is the BS that provides the  maximum received power averaged over fading. Thus 
\begin{align}
{\bf x}^*=\argmax_{{\bf x} \in \{{\bf x}^{*}_k\}} P_k \|{\bf x}\|^{-\alpha},
\end{align}
where ${\bf x}^*_k=\argmax_{{\bf x} \in {\Phi_k}} P_k \|{\bf x}\|^{-\alpha}$ is the location of the candidate serving BS in $\Phi_k$. 
In this letter, we are interested in a  fine-grained analysis of $\sir$ in terms of its meta distribution which is defined as follows. 
\begin{ndef}\label{def::meta::distribution}
The meta distribution of $\sir$  is  the CCDF of the conditional success probability $P_s(\beta) \triangleq\P(\sir>\beta|\Phi)$, i.e., 
\begin{equation}\label{eq::meta::dist::def}
\bar{F}(\beta,\theta) = \P[P_s(\beta) >\theta], \beta\in\R^+, \theta\in(0,1].
\end{equation}
\end{ndef}
Due to the ergodicity of $\Phi$, $\bar{F}(\beta,\theta)$ can be interpreted as the fraction of links in each realization of $\Phi$ that have an $\sir$ greater than $\beta$ with probability at least $\theta$.
According to Def.~\ref{def::meta::distribution}, the standard coverage probability~\cite{saha2019unified}  is the mean of $P_s(\theta)$ obtained by integrating \eqref{eq::meta::dist::def} over $\theta \in [0,1]$.  
\section{Meta Distribution of $\sir$}\label{sec::meta::distributions}
In this section, we will construct the equivalent single tier representation  of the $K+1$ tier network defined in Section~\ref{sec::system::model} by projecting $\Phi\subset\R^2$ on $\R^+$.  
For  a PPP-based model, the equivalent network in $\R^+$ remains analytically tractable~\cite{MadhusudhananRestrepoBrown2016} because of the application of the mapping theorem~\cite{chiu2013stochastic}, which is stated as follows. 
\begin{theorem}\label{thm::mapping}
If $\Phi$ is a PPP in $\R^d$ with intensity  $\lambda(x)$ and $f:\R^d\mapsto \R^s$ is a measurable map with $\Lambda(f^{-1}\{y\})=0, \forall y\in\R^s$, then $f(\Phi) = \cup_{x\in\Phi}\{f(x)\}$ is a PPP with intensity measure   $\tilde{\Lambda}(B')= \int_{f^{-1}(B')} \lambda(x){\rm d}{x}$, for all compact $B'\in\R^s$. 
\end{theorem}
Following Thm.~\ref{thm::mapping},  since $\{{\bf z}\}=\Phi_{{\rm p}_i}\subset\R^2$ ($\forall i\in {\cal K}_2$) is a homogeneous PPP with intensity $\lambda_{{\rm p}_i}$, then $\tilde{\Phi}_{{\rm p}_i}\triangleq \{\|{\bf z}\|\}$ is an inhomogeneous PPP in $\R^{+}$ with intensity and density: 
\begin{equation}
\tilde{\lambda}_{{\rm p}_i}(z)= 2\pi\lambda_{{\rm p}_i}z,\ \tilde{\Lambda}_{{\rm p}_i}(z) = \pi\lambda_{{\rm p}_i}z^2, z>0,
\end{equation}
respectively. Since Theorem~\ref{thm::mapping} does not hold when $\Phi_{i}$ ($i\in{\cal K}_2$) is PCP, the projection of $\Phi$ on $\R^+$ cannot be handled on similar lines of ~\cite{MadhusudhananRestrepoBrown2016}. 
The key enabler of our analysis is the following property of PCP which allows the application of Theorem~\ref{thm::mapping} to $\Phi_i$ for $i\in {\cal K}_2$. 
\begin{lemma} \label{lem: PCP and PPP}
If  $\Phi_i\equiv\{{\bf x}\}\subset\R^2$ is a PCP, then the sequence $\bar{\Phi}_i\triangleq\{\|{\bf x}\|\}\subset \R^+$ conditioned on  $\tilde{\Phi}_{{\rm p}_i}$  is  an  inhomogeneous PPP  with density and intensity: 
\begin{align}\label{eq: conditional density of i-th}
&\bar{\Lambda}_i({x}|\tilde{\Phi}_{{\rm p}_i})= \bar{m}_i \sum_{ z  \in \tilde{\Phi}_{{\rm p}_i}} F_{{\rm d}_i}(x|z), x>0;\quad  \forall i\in {\cal K}_2,\notag\\
&\bar{\lambda}_i({x}|\tilde{\Phi}_{{\rm p}_i})=  \bar{m}_i \sum_{ z  \in \tilde{\Phi}_{{\rm p}_i}}f_{{\rm d}_i}(x|z).
\end{align}
%with  $ \bar{G}_i\Big(({x} {P_i})^{1/\alpha} |z\Big)=$
%\begin{align}
%&  \int_0^{({x}{P_i})^{1/\alpha}}   f_{{\rm d}_i} (x'|z) x' {\rm d} x', \\
%&g_i\Big(({x} {P_i})^{1/\alpha} |z\Big)=\frac{\rm d}{{\rm d} {x}} \bar{G}_i\Big(({x} {P_i})^{1/\alpha} |z\Big),
%\end{align}
\end{lemma} 
\begin{IEEEproof} See {\cite[Prop.~1]{saha2019unified}.} %\chb{need to expand the proof.}
\end{IEEEproof}
Following the same argument, for \case~2 users, $\tilde{\Phi}_0 = \{\|{\bf z}_0\|\equiv z_0\}$ is also a PPP conditioned on $z_0$ with intensity $\bar{m}_0f_{{\rm d}_0}(x|z_0) \equiv \bar{m}_qf_{{\rm d}_q}(x|z_0)$.
Hence, we begin our  analysis by first conditioning on the locations of the points in every parent PPP, i.e., $\tilde{\Phi}_{{\rm p}_i}$, $\forall  i \in{\cal K}_2' \triangleq {\cal K}_2\cup \{0\}$.  Following Lemma~\ref{lem: PCP and PPP}, we have a sequence of BS PPPs $\{\tilde{\Phi}_{i},i\in{\cal K}_2'\}$ in $\R^+$ conditioned on $\cup_{i\in{\cal K}_2'}\tilde{\Phi}_{{\rm p}_i}$. 

Let us define $\tilde{\Phi}_i = \{P_i^{-1}\|{\bf x}\|^{\alpha},\forall {\bf x}\in\Phi_i\}$ as the projection of $\Phi_i$ ($\forall\ i \in{\cal K}_1\cup {\cal K}_2'$) on $\R^+$.  For $i\in{\cal K}_1$, using Theorem~\ref{thm::mapping}, the {density} of this 1-D inhomogeneous PPP $\tilde{\Phi}_i$ is
\begin{align*}
\tilde{\Lambda}_i({x})=  \int_0^{2\pi} \int_0^{({x}{P_i})^{1/\alpha}}  \lambda_i  x' {\rm d} x' {\rm d} \theta= \pi \lambda_i ({x}{P_i})^{\frac{2}{\alpha}}.
\end{align*}
For $\Phi_i (i \in {\cal K}_2')$ conditioned on $\tilde{\Phi}_{{\rm p}_i}$, the {density} of $\tilde{\Phi}_i = \{P_i^{-1}x^\alpha, \forall x\in\bar{\Phi}_i\}$ is: 
% $\tilde{\Lambda}_i({x}|\tilde{\Phi}_{{\rm p}_i})=$
\begin{equation*}
 \tilde{\Lambda}_i({x}|\tilde{\Phi}_{{\rm p}_i})=\int_0^{({x}{P_i})^{\frac{1}{\alpha}}} \bar{m}_i \sum_{{z}\in \tilde{\Phi}_{{\rm p}_i}}  f_{{\rm d}_i} (x'|z) {\rm d} x'\\= \bar{m}_i \sum_{z\in \tilde{\Phi}_{{\rm p}_i}} F_{{\rm d}_i}\big((xP_i)^{\frac{1}{\alpha}}|z\big).
\end{equation*}
Using the superposition theorem for PPP~\cite{chiu2013stochastic}, the {density} function of the 1-D PPP $\tilde{\Phi}=\cup_{i\in {\cal K}} \tilde{\Phi}_i$, which is the  projection  of the  $K+1$ tier HetNet $\Phi$,  can be obtained as follows: 
\begin{align}
\tilde{\Lambda}({x}|\cup_{{i \in {\cal K}_2'}}\tilde{\Phi}_{{\rm p}_i} )&= \sum_{i \in {\cal K}_1}\Lambda_i({x}) +\sum_{i \in {\cal K}_2'}\Lambda_i({x}|\tilde{\Phi}_{{\rm p}_i})\notag\\
&=\sum\limits_{i\in{\cal K}_1}\pi\lambda_i(xP_i)^{\frac{2}{\alpha}} + \sum\limits_{i\in{\cal K}_2'}\bar{m}_i\sum\limits_{z\in\tilde{\Phi}_{{\rm p}_i}} F_{{\rm d}_i} \big((xP_i)^{\frac{1}{\alpha}}|z\big),\label{eq::density::PCP}
\end{align}
and intensity: 
\begin{align}
\tilde{\lambda}(x|\cup_{{i \in {\cal K}_2'}}\tilde{\Phi}_{{\rm p}_i})  &= \frac{{\rm d}}{{\rm d}x}\tilde{\Lambda}({x}|\cup_{{i \in {\cal K}_2'}}\Phi_{{\rm p}_i} )\notag\\
&=
\sum\limits_{i\in{\cal K}_1}\pi\lambda_i\frac{2}{\alpha}P_i^{\frac{2}{\alpha}}x^{\frac{2}{\alpha}-1} + \sum\limits_{i\in{\cal K}_2'}\bar{m}_i\sum\limits_{z\in\tilde{\Phi}_{{\rm p}_i}} P_i^{\frac{1}{\alpha}}\frac{1}{\alpha}x^{\frac{1}{\alpha}-1}  f_{{\rm d}_i} \big((xP_i)^{\frac{1}{\alpha}}|z\big).\label{eq::intensity::PCP}
\end{align}
We are now in a position to define  $\sir$ of the typical user  as:
\begin{equation}\label{eq::sir::1-D}
   \sir = \frac{h_{\tilde{x}^*}({\tilde{x}^*})^{-1}}{\sum_{\tilde{x}\in{\tilde{\Phi}},{x}>{\tilde{x}}^*}   h_{{x}} {x}^{-1} },
\end{equation}
where $\tilde{x}^* = \arg\min_{x\in\tilde{\Phi}} x$ is the point in $\tilde{\Phi}$ closest to the origin.  The equivalence of the $\sir$-s expressed in terms of $\Phi$ and $\tilde{\Phi}$ is formally stated in the following proposition.
\begin{prop}\label{rem::equivalence}
 The $\sir$ of a typical user in the $K+1$-tier HetNet  $\Phi\subset\R^2$ with the BSs of the $i$-th tier transmitting at power $P_i$ and $\max$-power based user association (defined in \eqref{eq::sir::def})  has the same distribution as that of a single tier 1-D network  $\tilde{\Phi}\subset \R^+$ with nearest BS association, where $\tilde{\Phi}$ is an inhomogeneous PPP with intensity  $\tilde{\lambda}({x}|\cup_{{i \in {\cal K}_2 \cup \{0\}}}\tilde{\Phi}_{{\rm p}_i} )$, or equivalently, density $\tilde{\Lambda}({x}|\cup_{{i \in {\cal K}_2 \cup \{0\}}}\tilde{\Phi}_{{\rm p}_i} )$  with 
 all BSs transmitting at unit power. Here $\tilde{\Phi}_{{\rm p}_0}=\emptyset$ for \case~1 users and $\tilde{\Phi}_{{\rm p}_0}={z}_0\sim f_{{\rm d}_q}( {z}_0|0)$ for \case~2 users. 
 \end{prop}
 The distribution of $\tilde{\Phi}$ being unknown, the main contribution of the paper is to use the fact that conditional version of $\tilde{\Phi}$ given $\cup_{i\in{\cal K}_2'}\tilde{\Phi}_{{\rm p}_i}$ is a PPP. We will then leverage the tractability of the PPP under the conditioning of the parent PPs and decondition with respect to $\cup_{i\in{\cal K}_2'}\tilde{\Phi}_{{\rm p}_i}$ at the last step of the analysis. 
 Since $\tilde{\Phi}|\cup_{i\in{\cal K}_2'}\tilde{\Phi}_{{\rm p}_i}$ is a PPP, the PDF of $\tilde{x}^{*}$ is given by~\cite{chiu2013stochastic}:
\begin{equation}\label{eq::contact::distance::PDF}
f_{\tilde{x}^{*}}(x)=  \lambda({x}|\cup_{i\in{\cal K}_2'}\tilde{\Phi}_{{\rm p}_i})\exp\big(-\Lambda(x|\cup_{i\in{\cal K}_2'}\tilde{\Phi}_{{\rm p}_i})\big), x>0.
\end{equation}
The direct calculation of the meta distributions being infeasible even for the baseline PPP-based models~\cite{wang2018sir},  we first derive the expressions of the $b$-th order moments of $P_s(\beta)$: $M_b(\beta) \triangleq\E[P_s(\beta)^b]$.  Note that the coverage probability of the typical user in this setting, which was studied in our previous work~\cite{saha2019unified}, is a special case of this result and can be obtained directly by setting $b=1$.
\begin{theorem}\label{thm::b-th::moment}
The $b$-th moment of $P_s(\beta)$, $b\in\mathbb{C}$  can be expressed as: 
\begin{equation}\label{eq::b-th::moment}
M_b(\beta)=\sum\limits_{i\in{\cal K}_1}\pi\lambda_i\frac{2}{\alpha}P_i^{\frac{2}{\alpha}}\int\limits_{0}^{\infty}Q(r)\prod\limits_{j\in{\cal K}_2'}{\cal PG}_{{\tilde{\Phi}}_{{\rm p}_j}}(r)r^{\frac{2}{\alpha}-1}{\rm  d}r+
\sum\limits_{i\in{\cal K}_2'}\int\limits_{0}^{\infty}Q(r)\prod\limits_{j\in{\cal K}_2'\setminus\{ i\}}{\cal PG}_{\tilde{\Phi}_{{\rm p}_j}}(r){\cal SP}_{\tilde{\Phi}_{{\rm p}_i}}(r){\rm d}r,
\end{equation}
where  
\begin{align}\label{eq::Q}
Q(r) = \exp\bigg(-\pi r^{\frac{2}{\alpha}}\sum_{j\in{\cal K}_1}   \lambda_jP_j^{\frac{2}{\alpha}} {}_{2}{\cal F}_{1}\bigg(b,-\frac{2}{\alpha},\frac{-2+\alpha}{\alpha},-\beta\bigg)  \bigg),
\end{align}
where ${}_{2}{\cal F}_{1}$ is the hypergeometric function and 
\begin{subequations}\label{eq:PGSP}
\begin{align}
&{\cal PG}_{\tilde{\Phi}_{{\rm p}_j}}(r) :=\E\bigg[\prod\limits_{z\in \tilde{\Phi}_{{\rm p}_j}}g_j(r,z)\bigg],
\label{eq::PG}\\
\text{and }&{\cal SP}_{\tilde{\Phi}_{{\rm p}_i}}(r):=\E\bigg[\sum\limits_{z\in\tilde{\Phi}_{{\rm p}_i}}\rho_i(r,z)\prod\limits_{z'\in\tilde{\Phi}_{{\rm p}_i}}g_i(r,z')
\bigg]
\label{eq::SP}
\end{align}
\end{subequations}
are the probability generating functional (PGFL) and sum-product functional (SPFL) of $\Phi_{{\rm p}_j}$ and $\Phi_{{\rm p}_i}$, respectively ($i,j\in{\cal K}_2'$) 
with
\begin{subequations}
\begin{align}
&g_j(r,z) = \exp\bigg(-\bar{m}_j P_j^{\frac{1}{\alpha}}\frac{1}{\alpha}r^{\frac{1}{\alpha}-1}\int\limits_{r}^{\infty} u(r,x) f_{{\rm d}_j}\big((xP_j)^{\frac{1}{\alpha}}|z\big){\rm d}x-\bar{m}_j F_{{\rm d}_j}\big((rP_j)^{\frac{1}{\alpha}}|z\big)\bigg),\label{eq::g_i}\\
&\rho_i(r,z) =\bar{m}_i P_i^{\frac{1}{\alpha}}\frac{1}{\alpha}r^{\frac{1}{\alpha}-1}f_{{\rm d}_i}\big((rP_i)^{\frac{1}{\alpha}}|z\big),\label{eq::rho}
\end{align}
\end{subequations}
where 
%\begin{equation}\label{eq::u}
$u(r,x) = (1-(1+{\beta r}/{x})^{-b}).$
%\end{equation}
\end{theorem}
\begin{IEEEproof} 
See Appendix~\ref{app::b-th::moment}.
\end{IEEEproof}
We note that $P_s(\theta)$ in \eqref{eq::b-th::moment} is expressed  in terms of  the PGFL and SPFL of $\tilde{\Phi}_{{\rm p}_i}$. Hence we are left with  deriving  the expressions of PGFL and SPFL of $\tilde{\Phi}_{{\rm p}_i}$ for $i\in{\cal K}_2'$. When $i\in{\cal K}_2$, the PGFL and SPFL of $\tilde{\Phi}_{{\rm p}_{i}}$ are known since it is a PPP~\cite[Lemmas~5,6]{saha2019unified}. For \case~2 users, the PGFL and SPFL of $\tilde{\Phi}_{{\rm p}_0}$ can be obtained by deconditioning over $z_0$, i.e. ${\cal PG}_{\tilde{\Phi}_{{\rm p}_0}}(r)=E_{z_0}[g_0(r,z_0)]$ and ${\cal SP}_{\tilde{\Phi}_{{\rm p}_0}}(r)=\E_{z_0}[\rho_0(r,z_0)g_0(r,z_0)]$.   We summerize the expressions of PGFL and SPFL in the following lemmas. 
\begin{lemma}\label{lemm::PGFL}
The PGFL of $\tilde{\Phi}_{{\rm p}_i}$ is given as: 
\begin{equation}\label{eq::PGFL}
{\cal PG}_{{\tilde{\Phi}_{{\rm p}_i}}}(r) = \begin{cases}
\exp\bigg(-\int\limits_{0}^{\infty}2\pi\lambda_{{\rm p}_i}z(1-g_i(r,z)){\rm d}z\bigg), &i\in{\cal K}_2,\\
\int\limits_{0}^{\infty}g_q(r,z) f_{{\rm d}_q}(z|0) {\rm d}z, &i=0, \text{\case~2}.
\end{cases}
\end{equation}
\end{lemma}
\begin{lemma}\label{lemm::SPFL}
The SPFL of $\tilde{\Phi}_{{\rm p}_i}$ is given as: 
\begin{align}\label{eq::SPFL}
{\cal SP}_{{\tilde{\Phi}_{{\rm p}_i}}}(r) =
\begin{cases}
 \int_{0}^{\infty}2\pi\lambda_{{\rm p}_i}z \rho_i(r,z)g_i(r,z){\rm d} {z} \exp \bigg(- \int_{0}^{\infty}2\pi\lambda_{{\rm p}_i}z'(1- g_i(r,z')){\rm d}{z}'\bigg), &i\in{\cal K}_2,\\
 \int\limits_{0}^{\infty} \rho_q(r,z)g_q(r,z)f_{{\rm d}_q}(z|0){\rm d}z, &i=0,\text{\case~2}.
 \end{cases}
\end{align}
\end{lemma}
The final expression of $M_b(\beta)$ is obtained by substituting ${\cal PG}_{\tilde{\Phi}_{{\rm p}_i}}(r)$ and ${\cal SP}_{\tilde{\Phi}_{{\rm p}_i}}(r)$ given by \eqref{eq::PGFL} and \eqref{eq::SPFL} to \eqref{eq::b-th::moment}. 
The accuracy of these expressions for a two tier network is verified with the Monte Carlo simulations in Figs.~\ref{fig::meta::distribution::mean} and \ref{fig::meta::distribution::variance}. Note that we can also derive $M_b(\beta)$ on similar lines of \cite{saha2019unified} by conditioning on the association to the BSs of the $i$-th tier ($i\in{\cal K}$). However, the single tier projection presented in this letter offers an alternate and more compact derivation of $M_b(\beta)$. We note that $M_1(\beta)$ and $M_2(\beta)$ for \case~1 and 2 users converge to  $M_1(\beta)$ and $M_2(\beta)$ for the baseline PPP model (i.e. where $\Phi_i$ (for all $i\in {\cal K}$) and $\Phi_{\rm u}$ are homogeneous PPPs) as $\sigma_2$ increases. This is because of the fact that the PCP weakly converges to  a homogeneous PPP as the cluster size  tends to infinity~\cite[Sec.~IV-B]{saha20173gpp}.  
Further, for the two tier network considered in Fig~\ref{fig::meta::distribution}, following \cite{afshang2018equi}, it is possible to show that $M_b(\beta), \forall\ b\in\mathbb{C}$ remains the same if $(\lambda_1,\lambda_{{\rm p}_2},\sigma_2)$ is replaced by $(\lambda_1/k^2,\lambda_{{\rm p}_2}/k^2,\sigma_2k)$ for $k>0$.  
\subsection{Approximation of Meta Distributions}
\label{subsec::beta::approx}
From the $b$-th moment of the conditional success probability, the meta distribution of $\sir$ can be obtained by using the Gil-Pelaez theorem as:
\begin{equation}
\bar{F}(\beta,\theta) = \frac{1}{2} + \frac{1}{\pi}\int\limits_{0}^{\infty} \frac{{\rm Im}(e^{-jt\log\theta}M_{jt}(\beta))}{t}{\rm d}t, 
\end{equation}
 where ${\rm Im}(z)$ denotes the imaginary part of $z\in\mathbb{C}$.  As it can be readily observed, the  expression  of the exact meta distribution is not computationally efficient since it requires   integration over the imaginary moments.  Hence, following the approach of \cite{metadistribution,Wang_meta_distribution}, we approximate $\bar{F}(\beta,\theta)$ with a beta-kernel $
\bar{F}(\beta,\theta) \approx 1- \frac{1}{B(\theta_1,\theta_2)}\int\limits_{0}^{\theta} t^{\theta_1-1}(1-t)^{\theta_2-1}{\rm d}t,
$%\end{equation}
 where $B(\cdot,\cdot)$ is the beta function and $(\theta_1,\theta_2)$ is given by solving the following system of equations:
\begin{equation*}
M_1 = \frac{\theta_1}{\theta_1+\theta_2} \text { and } M_2 = \frac{\theta_1^2}{(\theta_1+\theta_2)^2} \Big(\frac{\theta_2}{\theta_1(\theta_1+\theta_2+1)}+1\Big).
\end{equation*} 
\begin{figure}
\centering
\subfigure[Mean  of  meta distribution. 
\label{fig::meta::distribution::mean}]{\includegraphics[width=.5\linewidth,trim={1cm  
 .1cm 0.1cm .1cm},clip]{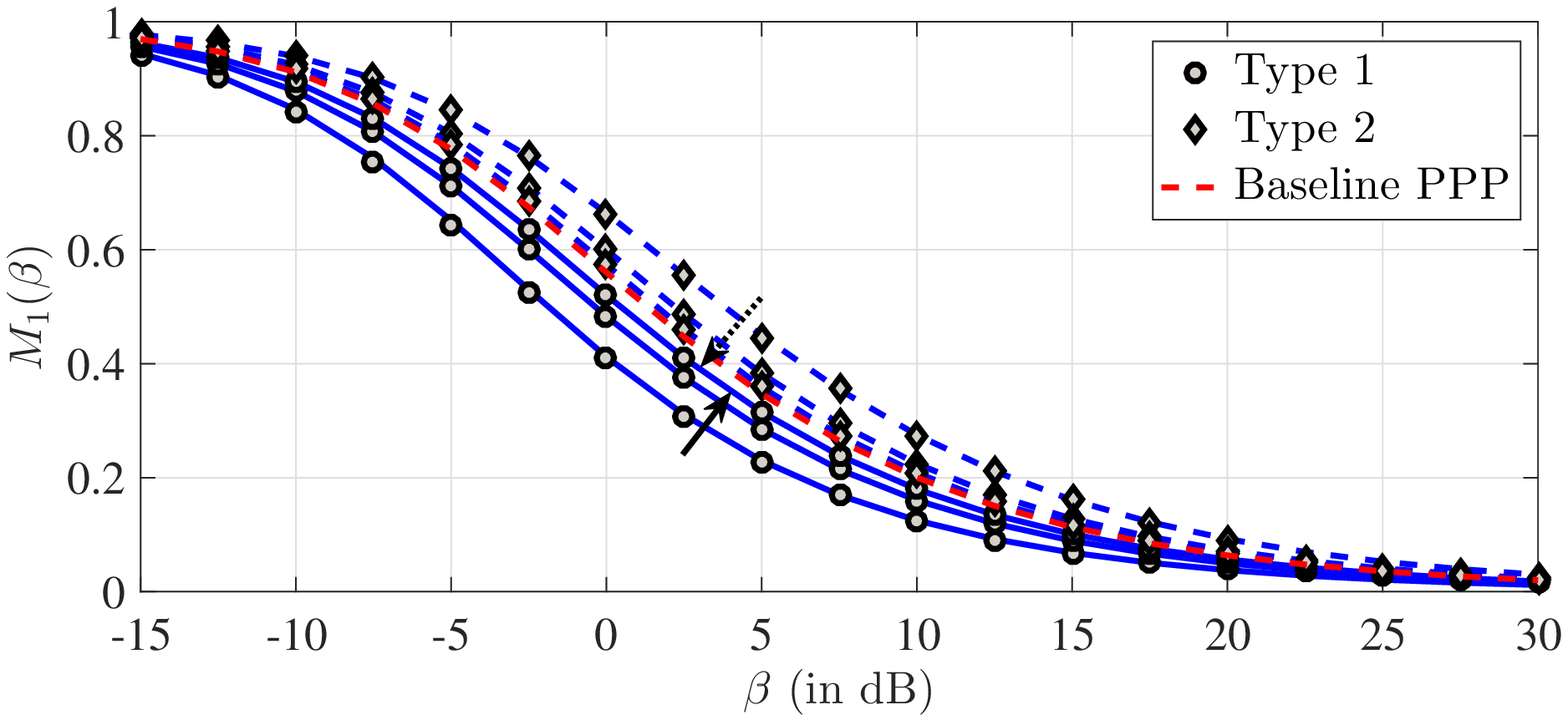}}%\vspace{-.15cm}
 \subfigure[Variance of  meta distribution. 
\label{fig::meta::distribution::variance}]{\includegraphics[width=.5\linewidth]{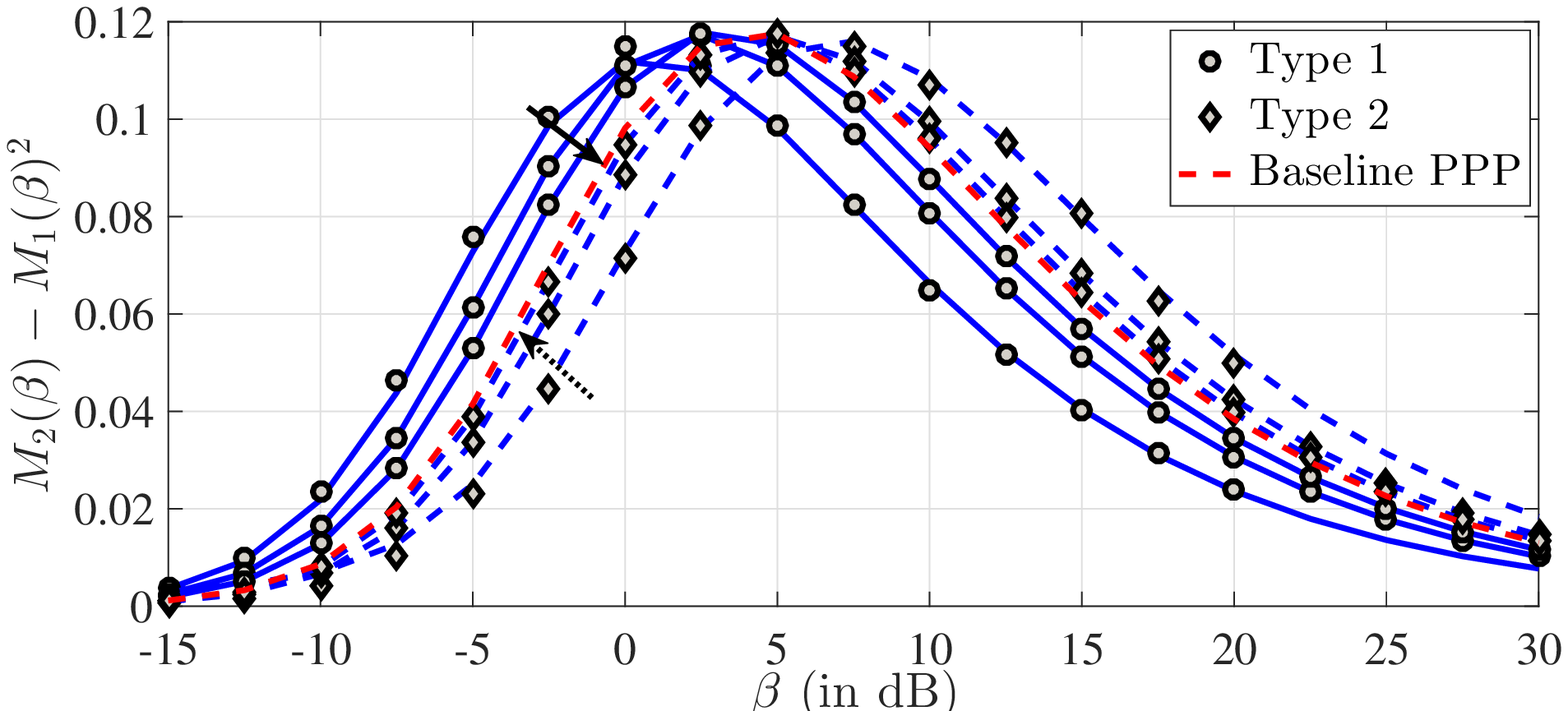}}%\vspace{-.15cm}
\\
\subfigure[Meta distribution. \label{fig::meta::distribution::beta}]{ \includegraphics[width=.45\linewidth,trim={1cm  
 .1cm 1cm .1cm},clip]{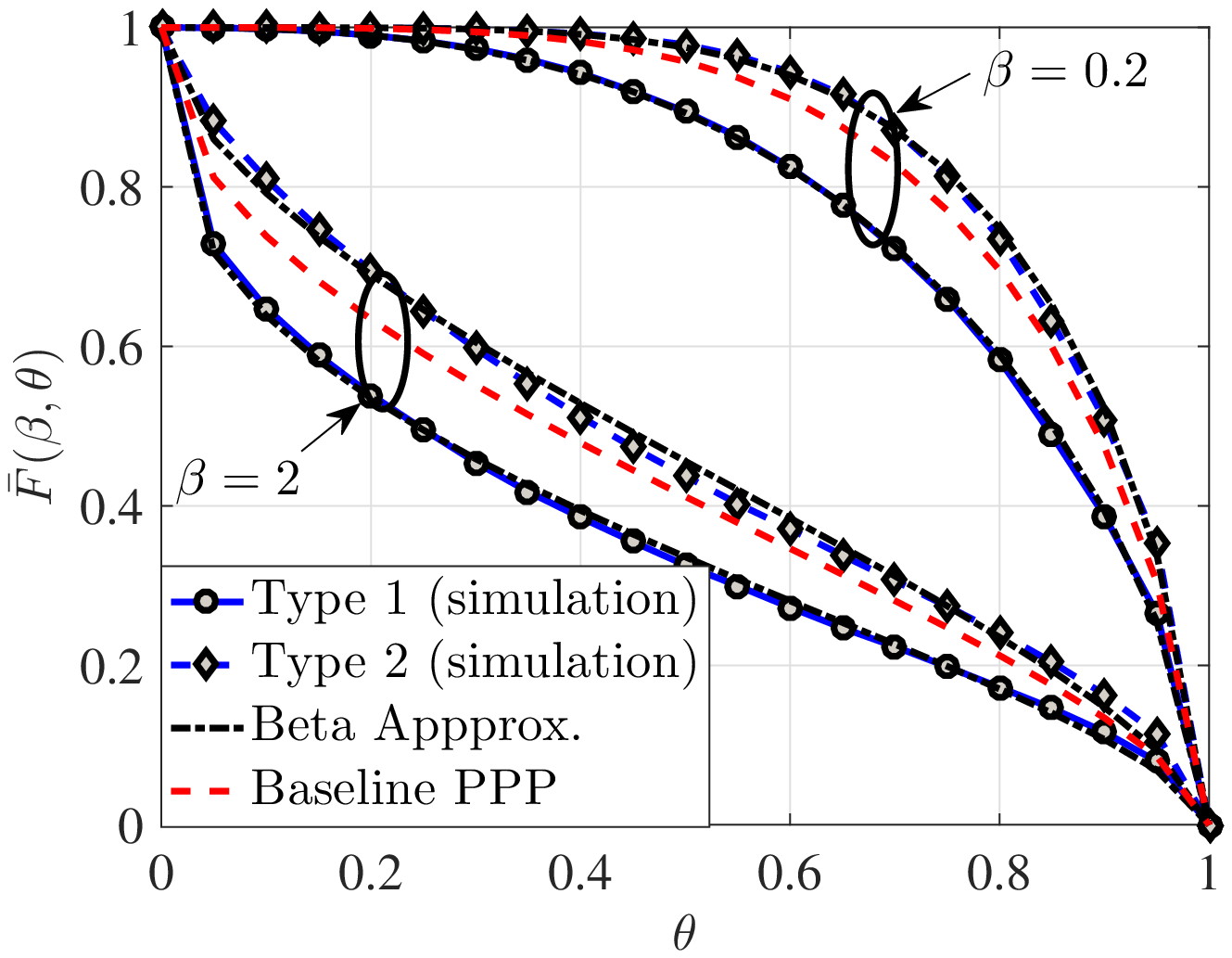}}
 \caption{Meta distribution of $\sir$ for \case~1 and \case~2 users in a two tier network. Details of the network configuration: $K=2$, ${\cal K}_1=\{1\}$, ${\cal K}_2=\{2\}$, $q=2$ for \case~2, 
 $\alpha=4$, $P_2 = 10^2P_1$, $\lambda_{{\rm p}_2} =2.5\ \text{km}^{-2}$, $\lambda_{{\rm p}_1} =1\ \text{km}^{-2}$,  $\bar{m}_2=4$, and $\sigma_2=\sigma_{\rm u}$. Markers indicate the values obtained from Monte Carlo simulations. The solid and dotted arrows in Fig.~\ref{fig::meta::distribution::mean} indicate the shift of the quantities with the increase in cluster size ($\sigma_2 = \{20,40,60\}$ m). For Fig.~\ref{fig::meta::distribution::beta}, $\sigma_2=40$ m.}\label{fig::meta::distribution}            
\end{figure}
In Fig.~\ref{fig::meta::distribution::beta}, we plot $\bar{F}(\beta,\theta)$ for a specific network configuration for {\sc Type}~1 and {\sc Type}~2 users. Clearly,  the beta approximation of $\bar{F}(\beta,\theta)$  is reasonably  tight  for a wide range of $\beta$. {Further, we observe that $\bar{F}(\beta,\theta)$ of \case~2 users is greater than $\bar{F}(\beta,\theta)$  in the baseline PPP-based model and $\bar{F}(\beta,\theta)$ of \case~1 users is less than $\bar{F}(\beta,\theta)$ in the baseline PPP-based model for all $\beta,\theta$}. 
This ordering is the same as the ordering observed for coverage probability (see Fig.~\ref{fig::meta::distribution::mean} and \cite[Sec.~IV]{saha2019unified}). However, it is a stronger result than the ordering of coverage. 
This implies that for a given $\beta$ and the same user density,  there exists more number of \case~2 users satisfying $\sir>\beta$ than \case~1 users in the network.  
% While Approaches 1 and 2 clearly differentiate the performance of \case~1 and \case~2 users in terms of the variation of coverage, their detailed analytical treatment is  out of scope of the manuscript whose primary objective is to develop the analytical framework for the exact characterization of the coverage probability. That said, in order to address this comment, we have included the insight obtained from the ordering of coverage probability for \case~1 and \Case~2 users discussed in the initial part of this response in Section~\ref{sec::results}. The new text is reproduced as follows.
\section{Conclusion}\label{sec::conclusion}
In this letter, we characterized the meta distribution of the
downlink $\sir$ for the typical user in the general $K$-tier HetNet model where the BSs are distributed as a PPP or a PCP.  The main technical
contribution is the accurate derivation of the $b$-th order moments of the conditional success probability. The key enabling step of the analysis is to condition on the parent point process of the BS PCPs which allows us to treat the PCPs as inhomogeneous PPPs. Under this conditioning, we obtain a sequence of BS PPPs in $\R^2$ which are projected on $\R^+$ to construct a single tier equivalence of the multi-tier HetNet. Using this single tier network, we present a compact derivation of the $b$-th order moments of the conditional success probability by applying the PGFL and SPFL of the parent PPPs. We finally use the moments of the conditional success probability to compute  a beta approximation of the meta distribution of $\sir$. 
\appendix
\subsection{Proof of Theorem~\ref{thm::b-th::moment}}
\label{app::b-th::moment}
From \eqref{eq::sir::1-D},  
\begin{align*}
P_s(\theta) &=\P\bigg(h_{\tilde{x}^*}>\beta \tilde{x}^*\sum_{{{{x}}\in{\tilde{\Phi}},x>\tilde{x}^*}}  h_{{x}} {x}^{-1}\bigg)\\&\stackrel{(a)}{=}
\E\bigg[\exp\bigg(-\beta \tilde{x}^*\sum\limits_{\substack{{{x}}\in{\tilde{\Phi}},\\x>\tilde{x}^*}}  h_{{x}} {x}^{-1}\bigg)\bigg] 
\\
&=\E\bigg[\prod\limits_{\substack{x\in\tilde{\Phi},x>\tilde{x}^*}}\big(1+\beta\tilde{x}^* x^{-1}\big)^{-1}\bigg].
\end{align*} 
Here $(a)$ follows from the CCDF of exponential distribution and the last step follows from the fact that $\{h_x\}$ is a sequence of i.i.d. exponential random variables.  
Now, 
\begin{align*}
P_s(\beta)^b|\cup_{i\in {\cal K}_2\cup\{0\}}\tilde{\Phi}_{{\rm p}_i}
&=\E\bigg[\prod
\limits_{x\in\tilde{\Phi},x>\tilde{x}^*}\big(1+\beta\frac{\tilde{x}^*}{x}\big)^{-b}\big|\bigcup\limits_{i\in{\cal K}_2'}\tilde{\Phi}_{{\rm p}_i}\bigg]\\
&\stackrel{(a)}{=} \exp\bigg(-\int\limits_{\tilde{x}^*}^{\infty} \underbrace{\bigg(1-{\bigg(1+\frac{\beta \tilde{x}^*}{x}}\bigg)^{-b}\bigg)}_{u(\tilde{x}^*,x)}\tilde{\lambda}(x|\cup_{i\in{\cal K}_2'}\tilde{\Phi}_{{\rm p}_i}){\rm d}x\bigg) \\
&\stackrel{(b)}{=}\int\limits_{0}^{\infty}  \exp\bigg(-\int\limits_{r}^{\infty} u(r,x)\tilde{\lambda}(x|\cup_{i\in{\cal K}_2'}\tilde{\Phi}_{{\rm p}_i}){\rm d}x\bigg) \exp\big(-\tilde{\Lambda}(r|\cup_{i\in{\cal K}_2'}\tilde{\Phi}_{{\rm p}_i})\big)
\tilde{\lambda}(r|\cup_{i\in{\cal K}_2'}\tilde{\Phi}_{{\rm p}_i}){\rm d}r \\
&= \int\limits_{0}^{\infty}  \exp\bigg(-\int\limits_{r}^{\infty} u(r,x)\tilde{\lambda}(x|\cup_{i\in{\cal K}_2'}\tilde{\Phi}_{{\rm p}_i}){\rm d}x-\tilde{\Lambda}(r|\cup_{i\in{\cal K}_2'}\tilde{\Phi}_{{\rm p}_i})\bigg) \bigg(
\sum\limits_{i\in{\cal K}_1}\pi\lambda_i\frac{2}{\alpha}P_i^{\frac{2}{\alpha}}r^{\frac{2}{\alpha}-1}\\
&\qquad\qquad+ \sum\limits_{i\in{\cal K}_2'}\bar{m}_i\sum\limits_{z\in\tilde{\Phi}_{{\rm p}_i}} P_i^{\frac{1}{\alpha}}\frac{1}{\alpha}r^{\frac{1}{\alpha}-1} f_{{\rm d}_i} \big((rP_i)^{\frac{1}{\alpha}}|z\big)\bigg) {\rm d}r\\
&\equiv {\cal T}_1\big(\cup_{i\in{\cal K}_2'}\tilde{\Phi}_{{\rm p}_i}\big) + {\cal T}_2\big(\cup_{i\in{\cal K}_2'}\tilde{\Phi}_{{\rm p}_i}\big),
\end{align*}
where 
\begin{align*}
{\cal T}_1\big(\cup_{i\in{\cal K}_2'}\tilde{\Phi}_{{\rm p}_i}\big) & = \sum\limits_{i\in{\cal K}_1}\pi\lambda_{i}\frac{2}{\alpha}P_i^{\frac{2}{\alpha}}\int\limits_{0}^{\infty} \exp\big(-\int\limits_{r}^{\infty}u(r,x)\tilde{\lambda}(x|\cup_{i\in{\cal K}_2'}\tilde{\Phi}_{{\rm p}_i}){\rm d}x-\tilde{\Lambda}(r|\cup_{i\in{\cal K}_2'}\tilde{\Phi}_{{\rm p}_i})\big)
    r^{\frac{2}{\alpha}-1}   {\rm d}r,\\
{\cal T}_2\big(\cup_{i\in{\cal K}_2'}\tilde{\Phi}_{{\rm p}_i}\big) & = 
\sum\limits_{i\in{\cal K}_2'}\bar{m}_i\int\limits_{0}^{\infty}\sum\limits_{z\in\tilde{\Phi}_{{\rm p}_i}} P_i^{\frac{1}{\alpha}}\frac{1}{\alpha}r^{\frac{1}{\alpha}-1} f_{{\rm d}_i} \big((rP_i)^{\frac{1}{\alpha}}|z\big)\\&\qquad\qquad\qquad\times \exp\big(-\int\limits_{r}^{\infty} {u(r,x)}\tilde{\lambda}(x|\cup_{i\in{\cal K}_2'}\tilde{\Phi}_{{\rm p}_i}){\rm d}x-\tilde{\Lambda}(r|\cup_{i\in{\cal K}_2'}\tilde{\Phi}_{{\rm p}_i})\big){\rm d}r.
\end{align*}
Here $(a)$ follows from the PGFL of the PPP (see Lemma~\ref{lemm::PGFL}), $(b)$ is obtained by deconditioning over $\tilde{x}^*$ whose PDF is given by~\eqref{eq::contact::distance::PDF}. 
 We are left with deconditioning ${\cal T}_1$ and ${\cal T}_2$ w.r.t. the distributions of the parent point processes for $i\in{\cal K}_2$ and $z_0$ for \case~2 users. We now derive the expressions of $\E[{\cal T}_1\big(\cup_{i\in{\cal K}_2'}\tilde{\Phi}_{{\rm p}_i}\big)]$ and $\E[{\cal T}_2\big(\cup_{i\in{\cal K}_2'}\tilde{\Phi}_{{\rm p}_i}\big)]$ as follows: 
\begin{align*}
 \E[{\cal T}_1\big(\cup_{i\in{\cal K}_2'}\tilde{\Phi}_{{\rm p}_i}\big)] &= \sum\limits_{i\in{\cal K}_1}\pi\lambda_{i}\frac{2}{\alpha}P_i^{\frac{2}{\alpha}}\int\limits_{0}^{\infty}\E\bigg[ \exp\bigg(-\int\limits_{r}^{\infty}u(r,x)\tilde{\lambda}(x|\cup_{i\in{\cal K}_2'}\tilde{\Phi}_{{\rm p}_i}){\rm d}x-\tilde{\Lambda}(r|\cup_{i\in{\cal K}_2'}\tilde{\Phi}_{{\rm p}_i})\bigg)\bigg]
    r^{\frac{2}{\alpha}-1}   {\rm d}r\\&= \sum\limits_{i\in{\cal K}_1}\pi\lambda_{i}\frac{2}{\alpha}P_i^{\frac{2}{\alpha}} r^{\frac{2}{\alpha}-1}  \int\limits_{0}^{\infty}\underbrace{\prod\limits_{j\in{\cal K}_1}\exp\bigg(-\int\limits_{r}^{\infty}u(r,x)\pi\lambda_j\frac{2}{\alpha}P_j^{\frac{2}{\alpha}}x^{\frac{2}{\alpha}-1}{\rm d}x
-\pi\lambda_j(rP_j)^{\frac{2}{\alpha}}\bigg)}_{Q(r)}\\&\qquad\qquad\times  \prod\limits_{j\in{\cal K}_2'}\E\bigg[\prod\limits_{z\in \tilde{\Phi}_{{\rm p}_j}}\underbrace{\substack{\exp\bigg(-\int\limits_{r}^{\infty}\bar{m}_j P_j^{\frac{1}{\alpha}}\frac{1}{\alpha}r^{\frac{1}{\alpha}-1} u(r,x)
f_{{\rm d}_j}\big((xP_j)^{\frac{1}{\alpha}}|z\big) {\rm d}x\bigg)\\-\bar{m}_j F_{{\rm d}_j}\big((rP_j)^{\frac{1}{\alpha}}|z\big)}}_{g_j(r,z)}\bigg] {\rm d}r,
\end{align*}
where the last step is obtained by substituting $\tilde{\Lambda}$ and $\tilde{\lambda}$ with \eqref{eq::density::PCP} and \eqref{eq::intensity::PCP}, respectively.  Applying the same substitution, 
\begin{align*}
\E\big[{\cal T}_2\big(\cup_{i\in{\cal K}_2'}\tilde{\Phi}_{{\rm p}_i}\big)\big]  &= 
\sum\limits_{i\in{\cal K}_2'}\int\limits_{0}^{\infty} \bar{m}_i   \E\bigg[\sum\limits_{z\in\tilde{\Phi}_{{\rm p}_i}} P_i^{\frac{1}{\alpha}}\frac{1}{\alpha}r^{\frac{1}{\alpha}-1} f_{{\rm d}_i} \big((rP_i)^{\frac{1}{\alpha}}|z\big)\\
&\qquad\qquad\times \exp\bigg(
  -\int\limits_{r}^{\infty} {u(r,x)}\tilde{\lambda}(x|\cup_{i\in{\cal K}_2'}\tilde{\Phi}_{{\rm p}_i}){\rm d}x-\tilde{\Lambda}(r|\cup_{i\in{\cal K}_2'}\tilde{\Phi}_{{\rm p}_i})
\bigg)\bigg]{\rm d}r\\&=\sum\limits_{i\in{\cal K}_2'}\int\limits_{0}^{\infty}\E\bigg[\sum\limits_{z\in\tilde{\Phi}_{{\rm p}_i}} \bar{m}_i P_i^{\frac{1}{\alpha}}\frac{1}{\alpha}r^{\frac{1}{\alpha}-1} f_{{\rm d}_i} \big((rP_i)^{\frac{1}{\alpha}}|z\big) 
    \exp\bigg(-\int\limits_{r}^{\infty} u(r,x)\bigg(\sum\limits_{j\in{\cal K}_1}\pi\lambda_j\frac{2}{\alpha}P_j^{\frac{2}{\alpha}}x^{\frac{2}{\alpha}-1} \\&\qquad\qquad+ \sum\limits_{j\in{\cal K}_2'}\sum\limits_{z\in\tilde{\Phi}_{{\rm p}_j}}\bar{m}_j 
P_j^{\frac{1}{\alpha}}\frac{1}{\alpha}x^{\frac{1}{\alpha}-1} 
f_{{\rm d}_j} \big((xP_j)^{\frac{1}{\alpha}}|z\big)\bigg){\rm d}x-\bigg(
\sum\limits_{j\in{\cal K}_1}\pi\lambda_j(rP_j)^{\frac{2}{\alpha}} 
\\&\qquad\qquad\qquad+ \sum\limits_{j\in{\cal K}_2'}\sum\limits_{z\in\tilde{\Phi}_{{\rm p}_j}} \bar{m}_j F_{{\rm d}_j} \big((rP_j)^{\frac{1}{\alpha}}|z\big)
\bigg)\bigg)\bigg]{\rm d}r\\
&=\sum\limits_{i\in{\cal K}_2'}\int\limits_{0}^{\infty}
\prod\limits_{j\in{\cal K}_1} \exp\bigg(-\int\limits_{r}^{\infty} {u(r,x)}\pi\lambda_j\frac{2}{\alpha}P_j^{\frac{2}{\alpha}}x^{\frac{2}{\alpha}-1}{\rm  d}x-\pi\lambda_j(rP_j)^{\frac{2}{\alpha}}\bigg)
\E\bigg[\sum\limits_{z\in\tilde{\Phi}_{{\rm p}_i}} \bar{m}_i P_i^{\frac{1}{\alpha}}\frac{1}{\alpha}r^{\frac{1}{\alpha}-1} \\
&\qquad\qquad\times  f_{{\rm d}_i} \big((rP_i)^{\frac{1}{\alpha}}|z\big) \prod\limits_{j\in{\cal K}_2'}\prod\limits_{z\in\tilde{\Phi}_{{\rm p}_j}}
\exp\bigg(- \int\limits_{r}^{\infty} u(r,x)\bar{m}_j P_j\frac{1}{\alpha}x^{\frac{1}{\alpha}-1} f_{{\rm d}_j} \big((xP_j)^{\frac{1}{\alpha}}|z\big){\rm d}x
\\&\qquad\qquad\qquad\qquad-\bar{m}_j F_{{\rm d}_j}\big((rP_j)^{\frac{1}{\alpha}}|z\big)
\bigg]{\rm d}r 
\\&=\sum\limits_{i\in{\cal K}_2'}\int\limits_{0}^{\infty}
Q(r)\prod\limits_{j\in{\cal K}_2'\setminus \{i\}}{\cal PG}_{\tilde{\Phi}_{{\rm p}_j}}(r) \E\bigg[\sum\limits_{z\in\tilde{\Phi}_{{\rm p}_i}}\underbrace{P_i^{\frac{1}{\alpha}}\frac{1}{\alpha}r^{\frac{1}{\alpha}-1}\bar{m}_{i}f_{{\rm d}_i}\big((rP_i)^{\frac{1}{\alpha}}|z\big)}_{\rho_{i}(r,z)}\\
&\qquad\times\prod\limits_{z\in\tilde{\Phi}_{{\rm p}_i}}\underbrace{\exp\bigg(- \int\limits_{r}^{\infty}u(r,x)  \bar{m}_i P_i^{\frac{1}{\alpha}}\frac{1}{\alpha}x^{\frac{1}{\alpha}-1} f_{{\rm d}_i} \big((xP_i)^{\frac{1}{\alpha}}|z\big){\rm d}x
-\bar{m}_iF_{{\rm d}_i}\big((rP_i)^{\frac{1}{\alpha}}|z\big)\bigg)
}_{g_i(r,z)}\bigg]{\rm d}r.
\end{align*}
The expression spanning over the last two lines can be identified as ${\cal SP}_{\tilde{\Phi}_{{\rm p}_i}}(r)$ (see \eqref{eq::SP}). 
In the above expressions, $Q(r)$ can be further simplified to \eqref{eq::Q}. 
%\chb{I think $Q(\tilde{r})$ can be obtained in closed- form format.  Evaluate it.}
%\chb{MA: It might be better to write ${\cal SP}$ and ${\cal PG}$ in terms of $\rho$  and $\mu$ }

\bibliographystyle{IEEEtran}
\bibliography{Letter_draft_arxiv.bbl}
\end{document}

%% file: notation.tex
\def\nb0{{\mathbf{0}}}
\def\nb1{{\mathbf{1}}}
\def\case{{\sc Type}}
\newtheorem{lemma}{Lemma}

\newtheorem{ndef}{Definition}

\newtheorem{theorem}{Theorem}
\newtheorem{prop}{Proposition}

\def\argmax{\operatorname{arg~max}}
\def\E{\mathbb{E}}

\def\P{\mathbb{P}}

\def\R{\mathbb{R}}

\def\sinr{\mathtt{SINR}}			% Signal to interference plus noise ratio

\def\sir{\mathtt{SIR}}

\def \x{\nu_0}